\documentstyle[preprint,aps,epsfig,axodraw]{revtex}
\newcommand{\no}     {\nonumber\\}
\newcommand{\ep}	{\epsilon}

\newcommand{\be}{\begin{equation}}
\newcommand{\ee}{\end{equation}}
\newcommand{\bea}{\begin{eqnarray}}
\newcommand{\eea}{\end{eqnarray}}
\newcommand{\Cdot}{\hspace{-1mm}\cdot\hspace{-1mm}}

\newcommand{\jpsi}{{J/\psi}}

\newcommand{\pol}{\rho_{\lambda\lambda^\prime}}

\newcommand{\la}{\lambda}
\newcommand{\sig}{\sigma}
\newcommand{\pr}{\prime}

\begin{document}
\draft
\preprint{
\begin{tabular}{l}
\hbox to\hsize{\hfill SNUTP  01-040}\\[-2mm]
\hbox to\hsize{\hfill DESY 01-194}\\[-2mm]
\hbox to\hsize{\hfill  November, 2001}\\[-3mm]
\end{tabular}
}

\title{
Polarization Effects in $B\to K_1(1270)+\jpsi$ Decays}

\author{
$^{(a)}$G. Kramer,
$^{(b)}$H. S. Song,
and $^{(b)}$Chaehyun Yu.
}

\address{
$^{(a)}$ II. Institut f\"{u}r Theoretische Physik, Universit\"{a}t Hamburg,
D-22761 Hamburg, Germany \\
$^{(b)}$ Center for Theoretical Physics and School of Physics,
     Seoul National University, \\Seoul 151-742, Korea 
}

\maketitle
\vspace{2cm}

\begin{abstract}
The joint angular distribution of the decay $B\to \jpsi V$, where $V$ is an
axial vector or vector resonance, followed by the subsequent decay processes 
of the $\jpsi$ and $V$ is calculated using the covariant density method. 
In particular, the case where $V$ is the axial vector meson $K_1(1270)$ which 
decays into $K\rho$ is considered as well as the case that $V$ is the vector
meson $K^\ast(890)$ which decays into $K\pi$.
\end{abstract}

\section{Introduction}

Recently the decays $B^0 \to K_1^0(1270)\jpsi$ and $B^+ \to K_1^+(1270)\jpsi$
have been observed in the Belle detector at the KEKB asymmetric $e^+e^-$
collider with significant branching ratios of the order of $10^{-3}$ \cite{k1}.
So these decays might be useful for CP violation studies similar to the decay
$B \to K^{*}(892)\jpsi$. The final state in the decay $B \to K_1 \jpsi$ is a
mixture of $CP=\pm 1$ eigenstates depending on the angular momentum of the
$K_1$ and $\jpsi$. For CP violation studies, the relative strengths of the two
eigenstates must be known. These can be determined from an analysis of the 
joint angular correlation of the decay products of $K_1 \to K\rho$ and 
$\jpsi \to l^+l^-$, where $l$ may be an electron or muon, respectively 
\cite{k2}. Since the branching fraction of the decay $K_1 \to K\rho$ is quite
large about 42\% \cite{k3} and the decay $\jpsi \to l^+l^-$ has a rather
distinct signature, which easily can be measured, it is to be expected that 
the full angular correlation will be measured as soon as a large data sample
has been collected at Belle and at BABAR. In fact, this has been achieved 
recently for the decay $B \to K^{*}\jpsi$ with $K^{*} \to K\pi$ and 
$\jpsi \to l^+l^-$ with high precision at BABAR \cite{k4} and Belle \cite{k5}
improving earlier measurements by the CLEO \cite{k6} and CDF \cite{k7}
collaborations.\\

The joint angular distribution of a $B$ meson decaying into the vector
(or axial vector) particles $V_1$ and $V_2$ with subsequent decays of $V_1$
and $V_2$, respectively, is described by three angles. In the original
derivation of the angular correlation for the decay $B \to K^{*}\jpsi$ with
$K^{*} \to K\pi $ and $\jpsi \to l^+l^-$ \cite{k8} the usual polar and
azimuthal angles for the decays of $K^{*}$ and $\jpsi$ were used, describing
the decay matrix elements of the decay products for $B$, $K^{*}$ and $\jpsi$ in
the familiar helicity basis \cite{k9}. The CP parity information, however,
is more easily extracted by introducing the $B$ decay amplitudes $A_0$, 
$A_{\parallel}$ and $A_{\perp}$ in the transversity basis \cite{k2}. In this
basis, using in the transversity system angles $\theta_{tr}$ and $\phi_{tr}$,
defined as polar and azimuthal angles of the $l^+$ in the $\jpsi$ rest 
frame, the angular correlation was presented in \cite{k10}. Following the
earlier work the angular correlation for the decay chain $B \to K_1\jpsi$,
$K_1 \to K\rho$ and $\jpsi \to l^+l^-$ can be derived as well either in the
helicity or the transversity basis.
Since the matrix elements of the two decay processes $B \to K^{*}\jpsi$ and
$B \to K_1\jpsi$ have the same Lorentz covariant form, although the invariant
form factors can be different in size and CP properties, it is more convenient
to apply the method of covariant density matrices \cite{k11} instead. Then it
is quite simple to incorporate the fact that the two subsequent decays
$K^{*} \to K\pi$ and $K_1 \to K\rho$ produce different contributions in the 
decay density matrix of the $\jpsi$.\\

In this short note we derive the angular distribution of the decay $\jpsi \to
l^+l^-$ in terms of the density matrix $\pol$. The explicit form of $\pol$ for 
the two cases, $B \to K_1(1270)\jpsi$ and $B \to K^{*}(892)\jpsi$, is 
calculated covariantly. With this result the full angular distribution for
both decays are derived in the transversity basis. It is clear that this 
method is quite general and can be applied to any $B$ decay into two spin-1
particles (either vector or axial vector) and their subsequent decays.
 
\section{Decay of $\jpsi \to l^+ l^-$}

Neglecting weak effects the process $\jpsi(q) \to l^+(q_1)l^-(q_2)$ can be 
described by the amplitude
\be
{\cal M} =\frac{f_1}{2} \ep^\mu(q) \bar{u}(q_2) \gamma_\mu v(q_1),
\ee
where $f_1$ is a coupling constant.
Then the decay distribution of the $\jpsi$ with polarization vector 
$\ep^\mu(q)$ after it is produced in any process is obtained from
\be
|{\cal M}|^2 = |f_1|^2 \langle \ep_\mu(q)\ep_\nu^\ast(q)\rangle
(q_1^\mu q_2^\nu+q_1^\nu q_2^\mu -\frac{q^2}{2} g^{\mu\nu}),
\label{square}
\ee
where the lepton polarizations have been summed. 
Here, the ensemble averaged 
value $\langle \ep_\mu(q)\ep_\nu^\ast(q)\rangle$ can be replaced by the 
covariant density matrix $\rho_{\mu\nu}$ of the $\jpsi$ which is obtained 
explicitly from the $B$ decay process into the $\jpsi$.
The most general form of $\rho^{\mu\nu}$ can be written as \cite{k11}
\be
\rho_{\mu\nu} = \frac{1}{3}(-g^{\mu\nu}+\frac{q_\mu q_\nu}{m_1^2})
-\frac{i}{2m_1}\ep_{\mu\nu\sig\tau}q^\sig   {\cal P}^\tau
-\frac{1}{2}{\cal Q}_{\mu\nu},
\label{covariant}
\ee
where the polarization vector ${\cal P}^\tau$
and the polarization tensor ${\cal Q}^{\mu\nu}$ can be obtained explicitly.
The covariant density matrix $\rho^{\mu\nu}$ is related to 
the density matrix $\rho_{\la\la^\pr}$ in the spin momentum basis as follows 
\be
\rho^{\mu\nu} = \ep^\mu(q,\la)\rho_{\la\la^\pr}\ep^{\nu\ast}(q,\la^\pr).
\ee
Here the projection operation $\ep^\mu(q,\la)\ep^{\nu\ast}(q,\la^\pr)$
is given by \cite{k11}
\be
\ep^\mu(q,\la)\ep^{\nu\ast}(q,\la^\pr)
=\frac{1}{3}\bigg(-g^{\mu\nu}+\frac{q^\mu q^\nu}{m_1^2}\bigg) 
\delta_{\la^\pr\la}
-\frac{i}{2m_1}\ep^{\mu\nu\sig\tau}q_\sig n_\tau^i (S^i)_{\la^\pr\la}
-\frac{1}{2}n_i^\mu n_j^\nu (S^{ij})_{\la^\pr\la},
\label{projection}
\ee
where the $( S^i )$ are the standard matrix representations 
of spin-1 angular momentum operators
and $( S^{ij}) $ are traceless and symmetric matrices 
defined as
\be 
S^{ij} = S^i S^j + S^j S^i -\frac{4}{4} \delta^{ij} I.
\ee
In Eq. (\ref{projection}), $(n_0^\mu, n_1^\mu, n_2^\mu, n_3^\mu) $
form a tetrad with $n_0^\mu=(E/m, \vec{q}/m)$ and 
$n_3^\mu$ is the Lorentz boost from $\hat{q}$ in the $\jpsi$ rest frame 
\cite{k11}. Then ${\cal P}^\mu$ and ${\cal Q}^{\mu\nu}$ in 
Eq. (\ref{covariant}) can be written as
\bea
{\cal P}^\mu &=& n_i^\mu {\rm Tr}(S^i \rho) \no
{\cal Q}^{\mu\nu} &=& n_i^\mu n_j^\nu {\rm Tr} (S^{ij}\rho).
\eea
Then Eq. (\ref{square}) becomes
\be
|{\cal M}|^2 \sim |f|^2 m_1^2\rho_{\la\la^\pr}
\bigg[ \frac{1}{3}\delta_{\la^\pr\la} +\frac{1}{m_1^2}
q_1\Cdot n^i q_1\Cdot n^j (S^{ij})_{\la^\pr \la}\bigg],
\ee
In the transversity basis, defined as in Fig. 1
in the $\jpsi$ rest frame, it is
\bea
d {\mit\Gamma} &\sim&  \Big[~ 1+\rho_{00} + (1-3\rho_{00}) \sin^2 \theta_{tr}
\cos^2\phi_{tr} \no
&&\hspace{0.65cm}+2 {\rm Re} (\rho_{1 \mbox{-}1}) 
(\sin^2\theta_{tr} \sin^2\phi_{tr}
-\cos^2 \theta_{tr})\no
&&\hspace{0.65cm}-2 {\rm Im} (\rho_{1 \mbox{-}1}) 
\sin^2\theta_{tr} \sin 2\phi_{tr}\no
&&\hspace{0.65cm}+\sqrt{2} {\rm Re} (\rho_{1 0}- \rho_{\mbox{-}1 1}) 
\sin 2\theta_{tr} \cos \phi_{tr}\no
&&\hspace{0.65cm}-\sqrt{2} {\rm Im} (\rho_{1 0}+ \rho_{\mbox{-}1 1}) 
\sin 2\theta_{tr} \sin \phi_{tr}~\Big],
\label{cccc}
\eea
where $(\theta_{tr}, \phi_{tr})$ are the polar and azimuthal angles of the 
outgoing lepton $l^+$ in the transversity basis as defined in \cite{k2,k10}
where $(\hat{n}_3, \hat{n}_1, \hat{n}_2)$ 
is along the $(\hat{x}, \hat{y}, \hat{z})$
axis.

\section{Density matrix of $\jpsi$}

The explicit values of the density matrix elements
$\rho_{\la\la^\pr}$ are calculated from the amplitude of the 
$\jpsi$ production process. The matrix element of the
$B\to \jpsi V$ ($V$ is either $K_1$ or $K^\ast$) decay is given in terms of
three independent Lorentz scalars  $A$, $B$ and $C$ as \cite{k12,k8}
\be 
{\cal M} = A \ep_1^\ast\Cdot \ep_2^\ast
+\frac{B}{m_1m_2} \ep_1^\ast \Cdot q \ep_2^\ast \Cdot k
+\frac{iC}{m_1m_2} \ep^{\mu\nu\sig\tau} \ep_{1\mu}^\ast
\ep_{2\nu}^\ast q_\sig k_\tau,
\label{amplitude}
\ee
where $\ep_i (i =1,2)$ are the polarization vectors of $\jpsi$ and $V$, 
respectively, and $m_2$ and $k^\mu$ are the mass and momentum of the $V$ 
particle. Instead of the scalar amplitudes $a$, $b$ and $c$ 
it is more convenient to introduce the three transversity amplitudes $A_0, 
A_\parallel $ and $A_\perp$ if one wants to specify the $CP$ property of the 
decay process. They are related to $A, B, C$ by \cite{k10}
\bea
A_0 &=& -xA -(x^2-1)B, \no
A_\parallel &=& \sqrt{2} \hspace{1mm}A, \no
A_\perp &=& \sqrt{2(x^2-1)} \hspace{1mm}C,
\eea
where $x$ is defined as
\be
x= \frac{k \Cdot q}{m_1 m_2} = \frac{ m_B^2 - m_1^2 -m_2^2}{2m_1 m_2}.
\ee
When the decay of the $B$ into $\jpsi$ and $V$ is specified by the polarization 
vectors $\ep_1(q,\la_1)$ and $\ep_2(k,\la_2)$, respectively,
the density matrix elements of the $\jpsi$ and $V$ final states can be 
obtained from the square of Eq. (\ref{amplitude}) and 
by using Eq. (\ref{projection}) for each $\ep_i ( i= 1,2)$ as follows,
\bea
\rho_{\la_1\la_1^\pr, \la_2\la_2^\pr}&\sim&
\frac{1}{9} (|A_0|^2+|A_\parallel|^2+|A_\perp|^2)
\delta_{\la_1\la_1^\pr, \la_2\la_2^\pr} \no
&&-\frac{2}{3\sqrt{\mit\Delta}}{\rm Re} (A_\parallel^\ast A_\perp)
\{ m_1 n_1^i \Cdot k \hspace{1mm}( S^i)_{\la_1\la_1^\pr}
\delta_{\la_2\la_2^\pr} 
+ m_2 n_2^k \Cdot q \delta_{\la_1\la_1^\pr}
(S^k)_{\la_2\la_2^\pr}\} \no
&&+\frac{1}{\mit\Delta} 
\Big\{ \hspace{3mm}(|A_\parallel|^2+|A_\perp|^2)
m_1m_2 \hspace{1mm}n_1^i\Cdot k n_2^k\Cdot q \no
&&\hspace{1cm}+\sqrt{\frac{\mit\Delta}{2}} {\rm Im}(A_0^\ast A_\perp)
\langle q k n_1^i n_2^k\rangle \no
&&\hspace{1cm}-\frac{\sqrt{2}}{4}{\mit\Delta} {\rm Re}(A_0^\ast A_\parallel)
(n_1^i\Cdot n_2^k -\frac{4q\Cdot k}
{\mit\Delta} n_1^i\Cdot k n_2^k \Cdot q)\Big\}
(S^i)_{\la_1\la_1^\pr} (S^k)_{\la_2\la_2^\pr} \no
&&-\frac{1}{3{\mit\Delta}}
(2|A_0|^2-|A_\parallel|^2-|A_\perp|^2)
\Big\{ ~m_1^2\hspace{1mm} n_1^i\Cdot k \hspace{1mm}n_1^j\Cdot k\hspace{1mm} (S^{ij})_{\la_1\la_1^\pr}
\delta_{\la_2\la_2^\pr} \no
&&\hspace{5.35cm}+m_2^2\hspace{1mm} n_2^k\Cdot q\hspace{1mm} n_2^l\Cdot q\hspace{1mm} \delta_{\la_1\la_1^\pr}
(S^{kl})_{\la_2\la_2^\pr} \Big\} \no
&&-\frac{1}{\mit\Delta}  \Big\{
~\sqrt{2} {\rm Im} (A_0^\ast A_\parallel) \langle q k n_1^i n_2^k \rangle \no
&&\hspace{0.9cm} +\frac{2}{\sqrt{\mit\Delta}}
{\rm Re}(A_\parallel^\ast A_\perp) m_1 m_2 n_1^i\Cdot k n_2^k\Cdot q \no
&&\hspace{0.9cm}-\sqrt{\frac{\mit\Delta}{2}}
{\rm Re} (A_0^\ast A_\perp) 
\Big( n_1^i\Cdot n_2^k - \frac{4q\Cdot k}{\mit\Delta}
n_1^i\Cdot k n_2^k \Cdot q \Big) \Big\} \no
&&\hspace{5mm}\times \{ m_2 n_2^l\Cdot q\hspace{1mm} (S^i)_{\la_1 \la_1^\pr}
(S^{kl})_{\la_2\la_2^\pr} 
+ m_1 n_1^j\Cdot q\hspace{1mm} (S^{ij})_{\la_1 \la_1^\pr}
(S^{k})_{\la_2\la_2^\pr} \} \no
&&+\Big\{ \frac{4}{{\mit\Delta}^2}|A^0|^2 m_1^2 m_2^2
n_1^i \Cdot k n_1^j\Cdot k n_2^k \Cdot q n_2^l\Cdot q \no
&&\hspace{3mm} +\frac{1}{8} |A_\parallel|^2
( n_1^i\Cdot n_2^k -\frac{4q\Cdot k}{\mit\Delta}
  n_1^i \Cdot k n_2^k \Cdot q)
( n_1^j\Cdot n_2^l -\frac{4q\Cdot k}{\mit\Delta}
  n_1^j \Cdot k n_2^l \Cdot q) \no
&&\hspace{3mm} +\frac{1}{2{\mit\Delta}} |A_\perp|^2
\langle q k n_1^i n_2^k\rangle \langle q k n_1^j n_2^l\rangle \no
&&\hspace{3mm} -\frac{\sqrt{2}}{2{\mit\Delta}}{\rm Re}(A_0^\ast A_\parallel)
m_1 m_2 n_1^i \Cdot k n_2^k\Cdot q
( n_1^j\Cdot n_2^l -\frac{4q\Cdot k}{\mit\Delta}
  n_1^j \Cdot k n_2^l \Cdot q) \no
&&\hspace{3mm} +\frac{2\sqrt{2}}{{\mit\Delta} \sqrt{\mit\Delta}}
{\rm Im} (A_0^\ast A_\perp)
m_1 m_2 n_1^i \Cdot k n_2^k\Cdot q
\langle q k n_1^j n_2^l\rangle \no
&&\hspace{3mm}-\frac{1}{2 \sqrt{\mit\Delta}}
{\rm Im} (A_\parallel^\ast A_\perp)
( n_1^i\Cdot n_2^k -\frac{4q\Cdot k}{\mit\Delta}
  n_1^i \Cdot k n_2^k \Cdot q)
\langle q k n_1^j n_2^l\rangle  \Big\} 
(S^{ij})_{\la_1\la_1^\pr}
(S^{kl})_{\la_2\la_2^\pr}.
\label{dd}
\eea
Here the density matrix is multiplied  by 
$(|A_0|^2+|A_\parallel|^2+|A_\perp|^2)$ for convenience and 
${\mit\Delta}$ and $\langle q k n_1^j n_2^l\rangle $ are defined as
\bea
&&{\mit\Delta} = (m_B^4 +m_1^4 +m_2^4 -2m_B^2 m_1^2-2m_B^2 m_2^2-2m_1^2 m_2^2)
\no
&&\hspace{4.6mm}= 4[(q.k)^2 - m_1^2 m_2^2] = 4m_1^2 m_2^2 (x^2-1),
\label{delta} \\
&&\langle q k n_1^j n_2^l\rangle 
= \ep^{\mu\nu\sig\tau} q_\mu k_\nu (n_\sig^j)_1 (n_\tau^l)_2.
\eea

Now the density matrix element of $\jpsi$, after $V$ decays,
can be obtained by specifying the amplitude of the $V$ particle decay 
process.
If the explicit form of the $V$ decay process is given, 
one can explicitly obtain the density matrix of the $V$, 
$\rho_{\la_2^\pr \la_2}^D(V)$, for its decay process. Then the density matrix
of the $\jpsi$ can be obtained from the product
\be
\rho_{\la_1 \la_1^\pr, \la_2 \la_2^\pr}
\rho_{\la_2^\pr \la_2}^D(V).
\ee

The amplitude of the decay process
$K_1(k,\la_2)\to K(\tilde{k})\rho(\tilde{k}^\pr)$ is described by
\be
{\cal M}_2 =  a \ep_2^{} \Cdot \ep_3^{\ast}
+\frac{b}{m_2 m_\rho} \ep_2^{}\Cdot \tilde{k}^\pr \ep_3^{\ast} \Cdot k,
\label{amp}
\ee
where $\ep_3^\mu$ and $\tilde{k}^\pr$
are the polarization vector and momentum of $\rho$.
Equation (\ref{dd}) can be used for the decay process
$K_1\to K \rho$ 
if some notations are changed, and after
the polarization of  $\rho$ is summed over, we obtain
\be
\rho_{\la_2^\pr \la_2}^D(K_1) 
= \frac{1}{3}\delta_{\la_2^\pr \la_2} - \frac{2 m_2^2}{\tilde{\mit\Delta}}
(1-\xi)\hspace{1mm}
\tilde{k}\Cdot n_2^k \tilde{k}\Cdot n_2^l (S_2^{kl})_{\la_2^\pr \la_2} ,
\label{k1den}
\ee
where $\tilde{\mit\Delta}$ and $\xi$ are defined as 
\bea
\tilde{\mit\Delta} &=& (m_2^4 +m_K^4+m_\rho^4 -2 m_2^2 m_K^2 
-2 m_2^2 m_\rho^2-2 m_K^2 m_\rho^2), 
\label{tildedelta} \\
\xi &=& \frac{3 | a_\parallel|^2}{2(|a_0|^2+|a_\parallel|^2)}.
\eea
The linear polarization amplitudes in the $K_1$ decay are
related to the form factors in Eq. (\ref{amp}) and helicity amplitudes 
$H_0$ and $H_\pm$ as
\bea
a_0 &=& -\tilde{x} a - (\tilde{x}^2-1) b = H_0, \\
a_\parallel &=& \sqrt{2} a = ( H_{+1} +H_{-1})/\sqrt{2}=\sqrt{2} H_+, 
\eea
where $\tilde{x} = (m_2^2 -m_K^2+m_\rho^2)/2m_2 m_\rho$.

Likewise, the amplitude of the decay process
$K^\ast(k,\la_2) \to K(\tilde{k}) \pi(\tilde{k}^\pr)$
is given by
\be
{\cal M}_3 = f_3 \ep^\mu(k,\la_2)(\tilde{k}-\tilde{k}^\pr)_\mu
=2f_3\ep^\mu(k,\la_2) \tilde{k}_\mu,
\ee
and the density matrix of the decay process becomes
\be
\rho_{\la_2^\pr \la_2}^D(K^\ast) 
= 
\frac{1}{3}\delta_{\la_2^\pr \la_2} - \frac{2 m_2^2}{\tilde{\mit\Delta}}
\tilde{k}\Cdot n_2^k \tilde{k}\Cdot n_2^l (S_2^{kl})_{\la_2^\pr \la_2} ,
\label{kstarden}
\ee
where  $\tilde{\mit\Delta}$ is defined in the same way as in 
Eq. (\ref{tildedelta}) after replacing the subscripts $K, \rho$ by 
$K, \pi$.
We see that the density matrix for $K^\ast$ to $K \pi$ is obtained
from that of $K_1$ to $K \rho$ in Eq. (\ref{k1den}) if we replace 
$\xi=0$.

Considering the two processes simultaneously, we write
the decay density matrix elements of the two $V$ particles ($K_1$ and 
$K^\star $), Eqs. (\ref{k1den}) and (\ref{kstarden}), as
\be
\rho_{\la_2^\pr \la_2}^D = \frac{1}{3} \delta_{\la_2^\pr \la_2}
+D\tilde{k}\Cdot n_2^k \tilde{k}\Cdot n_2^l (S^{kl})_{\la_2^\pr \la_2},
\ee
and then we have 
\bea
D= -\frac{2m_2^2}{\tilde{\mit\Delta}},
\hspace{12.5mm}&&\hspace{5mm} {\rm for}
~K^\ast \to K\pi, \no
=-\frac{2m_2^2}{\tilde{\mit\Delta}}(1-\xi),&& \hspace{5mm} {\rm for}
~K_1 \to K\rho.
\eea

The explicit form of the density matrix elements of $\jpsi$ neglecting 
the subscript 1 in $\la_1$ and $n_1$ is now obtained as follows,
\bea
\rho_{\la\la^\pr} &\sim& \rho_{\la_1\la_1^\pr, \la_2\la_2^\pr}
                      \rho_{\la_2^\pr\la_2}^D \no
&=&\frac{1}{3} \delta_{\la\la^\pr}
\bigg[ ~\frac{1}{3} (|A_0|^2 +|A_\parallel|^2 +|A_\perp|^2)
\no
&&\hspace{1cm}+\frac{D\tilde{\mit\Delta}}{12m_2^2}
\Big\{1-\frac{48m_2^4}{{\mit\Delta} \tilde{\mit\Delta}} (q\Cdot J)^2\Big\}
(2|A_0|^2-|A_\parallel|^2-|A_\perp|^2)\bigg] \no
&& +(S^i)_{\la\la^\pr}
\bigg[ -\frac{2}{3\sqrt{\mit\Delta}} {\rm Re}(A_\parallel^\ast
A_\perp) m_1 k\Cdot n^i \no
&&\hspace{1.82cm} + \frac{2\sqrt{2}m_2}{{\mit\Delta}}
D\Big\{~2{\rm Im}(A_0^\ast A_\parallel) q\Cdot J
\langle q k \tilde{k} n^i \rangle \no
&&\hspace{3.9cm} +\frac{m_1}{24}\sqrt{2{\mit\Delta}}{\mit\Delta}
{\rm Re}(A_\parallel^\ast A_\perp)
\Big\{ 1-\frac{48m_2^4}{{\mit\Delta} \tilde{\mit\Delta}}
(q\Cdot J)^2\Big\}
 k\Cdot n^i \no
&&\hspace{3.9cm}+ \sqrt{\mit\Delta} {\rm Re}(A_0^\ast A_\perp) 
q\Cdot J (J\Cdot n^i -\frac{4 q\Cdot k q\Cdot J}{\mit\Delta}
k\Cdot n^i )\Big\}\bigg] \no
&&+(S^{ij})_{\la\la^\pr}
\bigg[ -\frac{m_1^2}{3{\mit\Delta}}
\bigg( 1+\frac{D \tilde{\mit\Delta}}{2m_2^2}\bigg)
(2|A_0|^2 -|A_\parallel|^2-|A_\perp|^2)
k\Cdot n^i k\Cdot n^j \no
&&\hspace{1.915cm}+D\bigg\{~
\frac{16m_1^2 m_2^2}{{\mit\Delta}^2}(q\Cdot J)^2
k\Cdot n^i k\Cdot n^j |A_0|^2 \no
&&\hspace{2.75cm}
+\frac{1}{2}|A_\parallel|^2 
(J\Cdot n^i - \frac{4 q\Cdot k q\Cdot J}{\mit\Delta} k\Cdot n^i )
(J\Cdot n^j - \frac{4 q\Cdot k q\Cdot J}{\mit\Delta} k\Cdot n^j ) \no
&&\hspace{2.75cm}
+\frac{2}{\mit\Delta}|A_\perp|^2
\langle q k \tilde{k} n^i \rangle
\langle q k \tilde{k} n^j \rangle \no
&&\hspace{2.75cm}
-\frac{4\sqrt{2}}{\mit\Delta} m_1 m_2 (q\Cdot J)
{\rm Re}(A_0^\ast A_\parallel)
k\Cdot n^i (J\Cdot n^j - \frac{4 q\Cdot k q\Cdot J}{\mit\Delta} k\Cdot n^j ) \no
&&\hspace{2.75cm}
-\frac{8\sqrt{2}}{\mit\Delta\sqrt{\mit\Delta}} m_1 m_2 (q\Cdot J)
{\rm Im} (A_0^\ast A_\perp)
k\Cdot n^i \langle q k \tilde{k} n^j \rangle \no
&&\hspace{2.75cm}
+\frac{2}{\sqrt{\mit\Delta}}
{\rm Im} (A_\parallel^\ast A_\perp)
(J\Cdot n^i - \frac{4 q\Cdot k q\Cdot J}{\mit\Delta} k\Cdot n^i )
\langle q k \tilde{k} n^j \rangle
\bigg\} \bigg],
\label{ccc}
\eea
where $J = -\tilde{k} + \frac{k \cdot \tilde{k}}{m_2^2}k$.
This density matrix is manifestly covariant and from the explicit
form of matrices $(S^i), (S^{ij})$,
one can obtain the angular distribution of
$B\to K_1\jpsi$ as well as that of $B\to K^\ast \jpsi$.
The angular distribution for the decay process 
$B\to K_1(1270) \jpsi~ (K_1\to K\rho, \jpsi \to l^+ l^-)$
in the transversity basis is derived from
Eqs. (\ref{cccc}) and (\ref{ccc}):
\bea
\frac{1}{\mit\Gamma}\frac{d {\mit\Gamma}}
{d \cos \theta_1 d \cos \theta_{tr} d \phi_{tr}}
&=&\frac{9}{32\pi}\frac{1}{1+3\alpha}\bigg[~ 
2 |A_0(t)|^2 \bigg\{ (\cos^2 \theta_1+\alpha)
(1-\sin^2\theta_{tr} \cos^2\phi_{tr})\bigg\} \nonumber \\
&&\hspace{1.23cm}
\hspace{1cm}+|A_\parallel(t)|^2\bigg\{\sin^2\theta_1(1-\sin^2\theta_{tr}
\sin^2\phi_{tr}) 
\nonumber \\
&&\hspace{1.23cm}
\hspace{2.7cm}+\alpha (1+\sin^2\theta_{tr} \cos^2\phi_{tr})\bigg\} \no
&&\hspace{1.23cm}\hspace{1cm}+|A_\perp(t)|^2\bigg\{\sin^2\theta_1\sin^2\theta_{tr} 
+\alpha (1+\sin^2\theta_{tr} \cos^2\phi_{tr})\bigg\} \no
&&\hspace{1.23cm}\hspace{1cm}+{\rm Im}\Big(A_\parallel^\ast(t)A_\perp(t)\Big)
\sin^2\theta_1 \sin 2\theta_{tr} \sin \phi_{tr} \nonumber \\
&&\hspace{1.23cm}\hspace{1cm}-\frac{1}{\sqrt{2}}{\rm Re}\Big(A_0^\ast(t)A_\parallel(t)\Big)
\sin 2\theta_1\sin^2\theta_{tr} \sin 2\phi_{tr} \nonumber \\
&&\hspace{1.23cm}\hspace{1cm}+\frac{1}{\sqrt{2}}{\rm Im}\Big(A_0^\ast(t)A_\perp(t)\Big)
\sin 2\theta_1\sin 2\theta_{tr} \cos \phi_{tr} ~\bigg],
\label{distribution}
\eea
where $\theta_1$ is the angle between the $K_1$ direction and the $K$ direction
in the $K_1$ rest frame
and the three amplitudes are normalized as 
$|A_0|^2+|A_\parallel|^2+|A_\perp|^2 =1$. 
Here, $\alpha$ is defined as
\be
\alpha = \frac{1}{3}\frac{\xi}{1-\xi} = \frac{|a_\parallel|^2}
{2|a_0|^2-|a_\parallel|^2}.
\ee
The angular distribution in the process $B\to  K^\ast\jpsi 
( K^\ast \to K \pi, \jpsi\to l^+l^-)$ in the transversity basis 
can be obtained by putting $\xi = 0$ or $\alpha=0$
in Eq. (\ref{distribution}).
The angular correlation for $B\to K_1 \jpsi$ depends also on
details of the decay $K_1\to K\rho$ in contrast to $K^\ast \to K \pi$
and this is related to the fact that the decay $K_1 \to K\rho$ depends on 
two independent matrix elements.
In order to analyze the transversity amplitudes for the $B$ decay,
one must know the ratio of three decay matrix elements
for $K_1\to K\rho$ or otherwise one must determine them from the angular
correlation together with the $B$ decay transversity amplitudes.

\acknowledgments

The authors would like to thank Prof. S. K. Kim for her suggestion to derive 
the new angular distribution. The work is supported 
by the DFG-KOSEF Collaboration (2000) under the contracts 446 KOR-113/137/0-1
(DFG) and 20005-111-02-2 (KOSEF). H. S. Song and G. Kramer thank their guest
institutes for the warm hospitality during visits in the summer and fall of 
2001.

\newpage
{\Large \bf Figure Captions}
\vskip 2cm

\begin{description}

\item
Fig. 1 :
The definitions of the angles in the transversity basis.
$\theta_{tr}$ and $\phi_{tr}$ are polar and azimuthal angles of $l^+$ 
in the $\jpsi$ rest frame.
$\theta_1$ is the angle between the $K_1$ direction and the $K$ direction
in the $K_1$ rest frame.

\end{description}

\begin{center}
\begin{figure}[htb]
\vspace{1cm}
\begin{picture}(400,400)
\Line(0,300)(300,50)
\Line(0,300)(100,400)
\Line(300,50)(400,150)
\Line(320,125)(80,325)
\Line(100,400)(254,271)
\Line(262,264)(295,237)
\Line(301,232)(400,150)
\LongArrow(290,150)(320,125)
\LongArrow(260,175)(260,330)
\LongArrow(260,175)(290,200)
\DashLine(260,175)(220,115){2}
\DashLine(310,250)(310,175){2}
\DashLine(260,175)(310,175){2}
\Curve{(260,210)(270,209)(280,202)}
\Curve{(280,175)(279,172)(273,162)}
\Curve{(128,285)(129,287)(138,290)}
\Text(330,250)[]{$l^+$}
\Text(290,165)[]{$\phi_{tr}$}
\Text(330,120)[]{$\hat{x}$}
\Text(300,200)[]{$\hat{y}$}
\Text(260,350)[]{$\hat{z}$}
\Text(270,220)[]{$\theta_{tr}$}
\Text(200,90)[]{$l^-$}
\Text(125,305)[]{$\theta_1$}
\Text(130,345)[]{$K$}
\Text(40,285)[]{$\rho$}
\Text(210,245)[]{$K_1$}
\Text(260,155)[]{$J/\psi$}
\LongArrow(170,380)(130,310)
\Text(175,385)[l]{in the $K_1$--rest frame}
\SetWidth{1.5}
\LongArrow(260,180)(146,275)
\LongArrow(220,115)(210,100)
\LongArrow(260,175)(310,250)
\LongArrow(140,275)(130,330)
\LongArrow(140,275)(50,285)
\end{picture}
\vspace{0.5cm}
\caption{\it 
}
\end{figure}
\label{fig1}
\end{center}

\end{document}